\documentclass[%
superscriptaddress,
amsmath,amssymb,
aip,
numerical,
]{revtex4-1}

\usepackage{graphicx}
\usepackage{subfig}
\graphicspath{{./figs/}}
\usepackage{dcolumn} 
\usepackage{bm} 
\usepackage{CJK}

\usepackage{changebar,xcolor,ulem}
\newif\ifdiff
\difffalse
\ifdiff
  \newcommand{\removed}[1]{\removedfragile{#1}}
  \newcommand{\removedfragile}[1]{{\color{red}{\sout{#1}}}{}}
  \newcommand{\added}[1]{\addedfragile{#1}}
  
  \newcommand{\changed}[2]{\removed{#1}\added{#2}}
\else
  \newcommand{\removed}[1]{} 
  \newcommand{\removedfragile}[1]{}
  \newcommand{\added}[1]{#1}
  
  \newcommand{\changed}[2]{\added{#2}}
\fi

\begin{document}

\preprint{APS/123-QED}

\title{Hydrodynamic turbulence in quasi-Keplerian rotating flows}

\author{Liang Shi}
\email{Email address: gliang.shi@gmail.com}
\affiliation{%
  Max Planck Institute for Dynamics and Self-Organization (MPIDS), 37077 G\"ottingen, Germany
}
\affiliation{Institute of Geophysics, University of G\"ottingen, 37077 G\"ottingen, Germany}
\author{Bj\"orn Hof}
\affiliation{%
  Max Planck Institute for Dynamics and Self-Organization (MPIDS), 37077 G\"ottingen, Germany
}
\affiliation{%
  Institute of Science and Technology Austria, 3400 Klosterneuburg, Austria
}

\author{Markus Rampp}
\affiliation{%
   Max Planck Computing and Data Facility, Gie{\ss}enbachstr.~2, 85748 Garching, Germany
}

\author{Marc Avila}
\email{Email address: marc.avila@zarm.uni-bremen.de}
\affiliation{%
 Institute of Fluid Mechanics, Friedrich-Alexander-Universit\"at Erlangen-N\"urnberg, 
 91058 Erlangen, Germany
}
\affiliation{%
  Center of Applied Space Technology and Microgravity, University of Bremen, 28359 Bremen, Germany
}

\date{\today}

\begin{abstract}
  We report a direct-numerical-simulation study of Taylor--Couette flow in
  the quasi-Keplerian regime at shear Reynolds numbers up to $\mathcal{O}(10^5)$. 
  Quasi-Keplerian rotating flow has been investigated for decades as a simplified 
  model system to study the origin of turbulence in accretion disks that is not fully understood. 
  The flow in this study is axially periodic and thus the experimental end-wall effects on 
  the stability of the flow are avoided. 
  Using optimal linear perturbations as initial conditions, 
  our simulations find no sustained turbulence: the strong initial perturbations distort 
  the velocity profile and trigger turbulence that eventually decays.
\end{abstract}

\maketitle

\section{Introduction}

In protoplanetary disks the inward accretion of matter is accompanied by an 
outward transport of angular momentum. In case of laminar flow  the 
momentum transport is solely governed by the fluid's molecular 
viscosity, $\nu$. The magnitude of the molecular viscosity is however much too small 
to account for the actually observed accretion rates. This discrepancy 
can be simply resolved by assuming that flows are turbulent which 
would considerably enhance the momentum transport.  
While the extremely large Reynolds numbers in such disks may be regarded as a 
justification for turbulence to occur, from a hydrodynamic stability 
perspective the situation is less clear. Disk flows have a Keplerian 
velocity profile with  $\Omega(r)\sim r^{-3/2}$, where $\Omega$ is 
the angular velocity. Such profiles are linearly stable according to 
the inviscid Rayleigh criterion~\cite{Rayleigh_prsla1917} and no purely 
hydrodynamic instability mechanism is known that would provide a 
direct path to turbulence. In hot ionized disks on the other hand 
turbulence can be triggered by the so-called magnetorotational 
instability~\cite{Balbus_aa1991,Balbus_revModPhys1998,Balbus_araa2003}, 
but this is thought to be of lesser importance in cold and 
weakly ionized disks. 
For the latter case alternative mechanisms have been suggested as potential 
sources of turbulence. Especially concerning density gradients several 
instabilities have been proposed in the literature (stratorotational 
instability~\cite{ShalybkovRuediger_aa2005,BarsGal_prl2007}; Zombie 
vortex instability~\cite{Marcus_apj2014}; 
Rossby wave instability~\cite{Lovelace_astroJ1999}; 
baroclinic instability~\cite{KlahrBodenheimer_apj2003}). Nevertheless, even in 
the absence of such instability mechanisms  turbulence could potentially 
arise from a nonlinear (subcritical) instability. Subcritical instabilities 
are for instance responsible for turbulence in pipe and related shear flows. 
Whether such a scenario is also responsible for turbulence in 
quasi-Keplerian rotating flows remains unclear.

This question has been recently studied in experiments of fluid flows 
between co-rotating cylinders, Taylor--Couette flow (TCf). By 
selecting appropriate rotation rates (corotation with a faster inner 
cylinder) velocity profiles can be established that have 
stability properties similar to Keplerian flows. Like in Keplerian flows, 
the angular velocity decreases outwards while the angular momentum 
increases and the flow is Rayleigh stable.  For this flow, Ji and 
co-workers~\cite{JiGoodman_nature2006,SchartmanGoodman_aa2012} have 
measured the Reynolds stress or the $\beta$ parameter introduced by 
Richard and Zahn~\cite{RichardZahn_aa1999} at discrete interior locations, 
and \added{at} Reynolds numbers (Re) up to $2\times 10^6$. They found that the experimentally 
measured $\beta$ is consistent with laminar flows and thus far below the 
value inferred from astrophysical observations. These authors concluded 
that hydrodynamic turbulence cannot account for the expected transport rate 
of angular momentum in disks. This was challenged by the experimental results 
of Paoletti \emph{et al}.~\cite{PaolettiLathrop_prl2011,PaolettiLathrop_aa2012}, 
who reported turbulent angular momentum transport in quasi-Keplerian TCf for 
Re above $10^5$.  Their estimated $\beta$ based on Torque measurements at the 
inner cylinder was found at similar level as in astrophysical disks. 
These contradictory conclusions are thought to arise because of design 
differences in the experiments, such as geometry (axial-length-to-gap aspect 
ratio $\Gamma$, and radius ratio $\eta$) and end-cap treatment as well as the 
measured physical quantities, making comparison difficult~\cite{Balbus_nature2011}. 

In the experiments of Ji and co-workers~\cite{JiGoodman_nature2006,SchartmanGoodman_aa2012}, 
the axial end walls were split into two independently rotating parts, 
whose rotation was selected as to minimize their effect on the bulk of the 
flow. The effectiveness of this strategy was demonstrated by  Obabko \emph{et al}.~\cite{ObabkoFischer_Phys2008}, who performed direct numerical simulation (DNS) of the same geometry and tested several different boundary conditions. In contrast, Paoletti \emph{et al}.~\cite{PaolettiLathrop_prl2011,PaolettiLathrop_aa2012} 
used a larger aspect ratio $\Gamma=11.47$ and measured the torque only around the mid-height of 
the experiment to avoid torque contributions arising near the end walls. 
However, their end walls were attached to the outer cylinder thereby generating
a very strong Ekman circulation, which was shown to entirely fill the apparatus unless $\Gamma\gg100$ were used~\cite{HollerbachFournier_aip2004,Edlund_pre2015}. 

Numerical simulations~\cite{Avila_prl2012} precisely reproducing the geometry and boundary conditions 
of the two aforementioned experimental setups~\cite{JiGoodman_nature2006,PaolettiLathrop_prl2011} showed that the axial end walls strongly disrupt quasi-Keplerian velocity profiles and cause turbulence to 
arise for Re as low as $\mathcal{O}(10^3)$. Although 
this explains why strong turbulence is found in the experiments of Paoletti 
\emph{et al.}~\cite{PaolettiLathrop_prl2011,PaolettiLathrop_aa2012}, 
as demonstrated later by the direct measurement of azimuthal velocity profiles 
performed by Nordsiek {\em et al.}~\cite{NordsiekLathrop_jfm2015}, it still appears 
to be in contradiction with the results of Ji and 
co-workers~\cite{JiGoodman_nature2006,SchartmanGoodman_aa2012}. However, 
similar measurements performed by Edlund and Ji~\cite{Edlund_pre2014} 
compellingly show that if the end wall boundary conditions are optimally chosen,
end-wall effects remain confined close to the axial boundaries and 
ideal laminar Couette profiles are obtained in the bulk of the experiments at sufficiently large Re. This 
was recently confirmed by direct numerical simulations of these experiments, which elucidated the progressive localization of turbulence at boundaries as Re increases up to 5$\times$10$^4$~\cite{LopezAvila_arxiv2016}.

Ostilla-M\'onico~\emph{et al.}~\cite{Monico_jfm2014} performed direct 
numerical simulations of TCf with axially periodic cylinders thereby 
eliminating end-wall effects. Their initial conditions were turbulent states 
obtained for stationary outer cylinder (Rayleigh-unstable regime) and 
at $t=0$ the rotation of the cylinders was suddenly changed to quasi-Keplerian 
(by impulsively increasing the rotation of the outer cylinder). Their simulations 
showed an immediate direct decay of turbulence in agreement with the 
experiments of Edlund and Ji~\cite{Edlund_pre2014}. Note 
however, that sudden changes in the driving velocity can also cause 
laminarization of flows that are turbulent if appropriately disturbed~\cite{Hof_Sci2010}. 
Further, while for stationary outer cylinder the dominant flow features are 
turbulent (toroidal) Taylor vortices rooted on the linear stability 
of the laminar flow~\cite{GrossmannSun_annRevFluidMech2016}, in quasi-Keplerian 
flows the disturbance with highest transient energy growth are (axially invariant) 
Taylor columns~\cite{Maretzke_jfm2014,Tuckerman_jfm2014}. These two issues 
raise the question of whether the initial conditions used by 
Ostilla-M\'onico~\emph{et al.}~\cite{Monico_jfm2014} and Lesur and 
Longaretti~\cite{LesurLongaretti_aa2005} are well suited as a trigger for 
turbulence in quasi-Keplerian flows. Following previous work on secondary 
instabilities~\cite{OrszagPatera_jfm1983,SchmidHenningson_springer2001,HoegbergHenningson_jfm1998} 
we perform direct numerical simulations of TCf with axially periodic 
cylinders starting from optimal perturbations superposed with very small 
three-dimensional random noise. Note that secondary means here that the laminar profile needs to be first disturbed with a ``primary'' disturbance so that  random noise can grow exponentially like in a linear instability. 
Our approach is also similar to the experiments 
of Edlund and Ji~\cite{Edlund_pre2014}, who apply strong injection disturbances 
to their quasi-Keplerian flow. Our simulations show transition to turbulence 
followed by its immediate decay at shear Reynolds number up to $10^5$. 

\section{Quasi-Keplerian Taylor--Couette flow}

\begin{figure}[!h]
  \centering
  \includegraphics[width=0.3\textwidth]{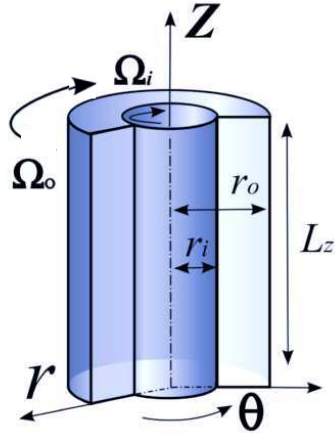}
  \caption{Sketch of the geometry of Taylor--Couette flow (TCf) in cylindrical
            coordinates. The inner and outer cylinders of radii $r_i$ and $r_o$ rotate independently
            at a speed of $\Omega_i$ and $\Omega_o$, respectively.  No-slip
            boundary conditions at the cylinders are used together with axially
            periodic boundary conditions. The fluid between the cylinders
            is driven by the shear force due to the molecular viscosity. }
  \label{fig:TCgeometry}
\end{figure}

Figure~\ref{fig:TCgeometry} shows a sketch of the geometry of TCf, the flow between two independently rotating concentric cylinders. The inner (outer) cylinder has radius $r_{i}$ $(r_o)$ and rotates at a speed of $\Omega_i$ $(\Omega_o)$. The Reynolds numbers of the inner and outer cylinder are defined as $Re_{i(o)}=\Omega_{i(o)}r_{i(o)}d/\nu$, where $d=r_o-r_i$ is the gap between the cylinders. The advective time unit,  $\tau_{d}=d/(r_i\Omega_i)$, based on the velocity of the inner cylinder is used in this paper. The geometry of TCf is fully specified by two dimensionless parameters: the radius-ratio $\eta=r_i/r_o$ and the length-to-gap aspect-ratio $\Gamma=L_z/d$, where $L_z$ is the axial length of the cylinders. The angular velocity of the laminar base flow, called circular Couette flow, is given by
\begin{equation}
  \begin{aligned}
    \Omega^b(r) &= C_1 + \frac{C_2}{r^2}, \\
    \text{ with } C_1=\frac{Re_o-\eta Re_i}{1+\eta},&\quad C_2=\frac{\eta(Re_i-\eta Re_o)}{(1-\eta)(1-\eta^2)},
  \end{aligned}
  \label{eq:baseProfile}
\end{equation}
which corresponds to a pure rotary shear flow.

\begin{figure}[!h]
  \centering
  \includegraphics[width=0.35\textwidth]{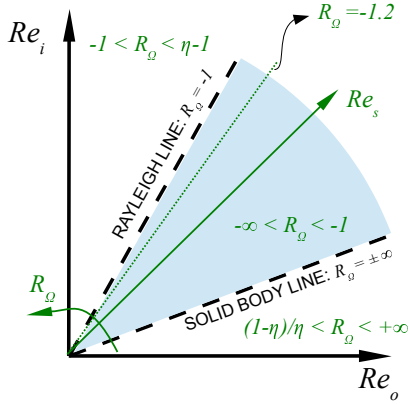}
  \caption[The parameter space $(Re_o,R_{i})$]{
    The parameter space ($Re_o,R_{i}$). The blue region represents the quasi-Keplerian regime. 
    The rotation number $R_{\Omega}$ is constant along  half-lines 
    starting from origin. The ranges of $R_{\Omega}$ are shown in different regions separated 
    by the Rayleigh line ($R_{\Omega}=-1$) and the solid body line ($R_{\Omega}=\pm\infty$). 
    The dotted line corresponds to the line $R_{\Omega}=-1.2$, on which our simulations are performed.}
  \label{fig:ResRoSpace}
\end{figure}

The dimensionless parameter choice introduced by Dubrulle {\em et al.}~\cite{DubrulleZahn_pof2005} is very useful as it separates rotation from shear
\begin{equation}
  \begin{aligned}
    Re_s &=\frac{2}{1+\eta}|\eta Re_o - Re_i|, \\
    R_{\Omega} &=\frac{(1-\eta)(Re_i+Re_o)}{\eta Re_o-Re_i}.
  \end{aligned}
  \label{eq:resro}
\end{equation}
The shear Reynolds number $Re_s$ characterizes the shear between the inner and outer cylinders and is essentially the square of the Taylor number, whereas the rotation number $R_{\Omega}$ is a measure for the mean rotation and is constant on every half-line out from the origin in the $(Re_o,Re_i)$-space (see Fig.~\ref{fig:ResRoSpace}). On the solid-body line, there is no relative motions between different layers and hence $Re_s=0$, whereas $R_{\Omega}=\pm \infty$. The quasi-Keplerian regime in TCf is the co-rotation region limited by the Rayleigh line and the solid-body line in the $(Re_o,Re_i)$ parameter space (the blue region in Fig.~\ref{fig:ResRoSpace}). The Rayleigh line ($Re_o=\eta Re_i$) separates linearly stable and unstable inviscid fluid flows. Below the Rayleigh line, the circular Couette flow is linearly stable. On the solid-body line, $\Omega_i=\Omega_o$ or $Re_i=\eta Re_o$, the fluids behave like a rigid body without shear, which means that all disturbances to the flow decay monotonically in time. In the quasi-Keplerian regime, the base velocity profiles satisfy two conditions: (1) radially increasing angular momentum ${d (\Omega^b(r) r^2)}/{dr}>0$; (2) radially decreasing angular velocity ${d\Omega^b(r)}/{dr}<0$.

\section{Numerical specification}

Our direct numerical simulations were performed at four different Reynolds numbers $Re_i=[1\times 10^4, 2\times 10^4, 1\times 10^5, 2\times 10^5]$ on the half line $R_{\Omega}=-1.2$, i.e., very close to the Rayleigh line $R_{\Omega}=-1$. This choice is motivated by Lesur and Longaretti~\cite{LesurLongaretti_aa2005}, who speculated that if there were a subcritical transition, this might be easier to trigger near the stability boundary. The corresponding shear Reynolds numbers are $Re_s=[5078.8, 10157.6, 50788, 101576]$. In order to compare with recent experimental and numerical results, the radius ratio is chosen to be $\eta = 0.71$. Another relevant parameter often used in the astrophysical literature is the local exponent of the angular velocity $q=-\text{dln}\Omega/\text{dln} r$. For a Keplerian 
velocity profile, $q=3/2$, and on the Rayleigh line $q=2$. Note that for circular Couette flow the parameter $q$ is not constant in the radial direction. In our simulations $q(r)=\frac{2C_2}{C_1r^2+C_2} \in [1.5,1.8]$, which is in the quasi-Keplerian regime. A brief comparison between astrophysical Keplerian flow and TCf of our simulations is shown in table~\ref{tab:KeplerianTCF}.

\begin{table} [!h]
  \centering
  \begin{tabular}{c c c c c c}
    \hline \hline 
     & $\Omega^b(r)$ & $Re_s$ & $R_{\Omega}$ & $q(r)$ & axial boundary \\
    \hline 
    TCf  &  $C_1r+C_2/r$ & $10^{4-5}$ & -1.2 & $[1.5,1.8]$ & periodic\\
    Keplerian &  $Cr^{-3/2}$ & $\gg 10^6$ & - 4/3 & 3/2 & free surfaces\\
    \hline \hline
  \end{tabular}
  \caption[Parameter comparison between astrophysical Keplerian flows and TCf]{
    A parameter comparison between TCf of our study and astrophysical Keplerian flows: 
    base angular velocity profile $\Omega^b(r)$, shear Reynolds number $Re_s$, rotation number 
    $R_{\Omega}$, local exponent $q(r)$ and axial boundary conditions. 
    $C_1$ and $C_2$ are defined in Eq.~\eqref{eq:baseProfile} while $C$ is another constant. }
  \label{tab:KeplerianTCF}
\end{table}

\begin{figure}[!ht]
  \centering
  \subfloat[$Re_i=2\times 10^4$]{\includegraphics[width=0.4\textwidth]{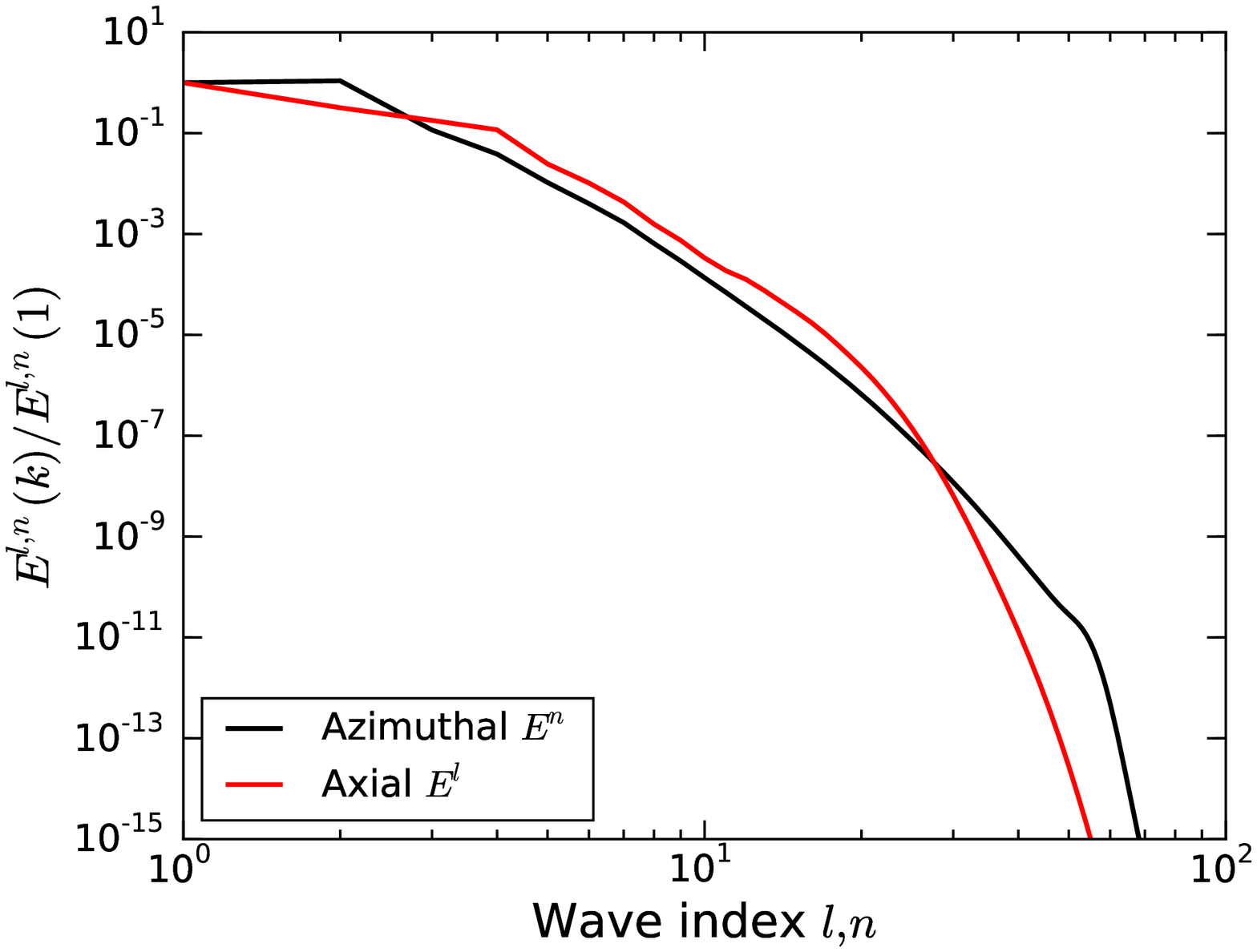}}
  \subfloat[$Re_i=2\times10^5$]{\includegraphics[width=0.4\textwidth]{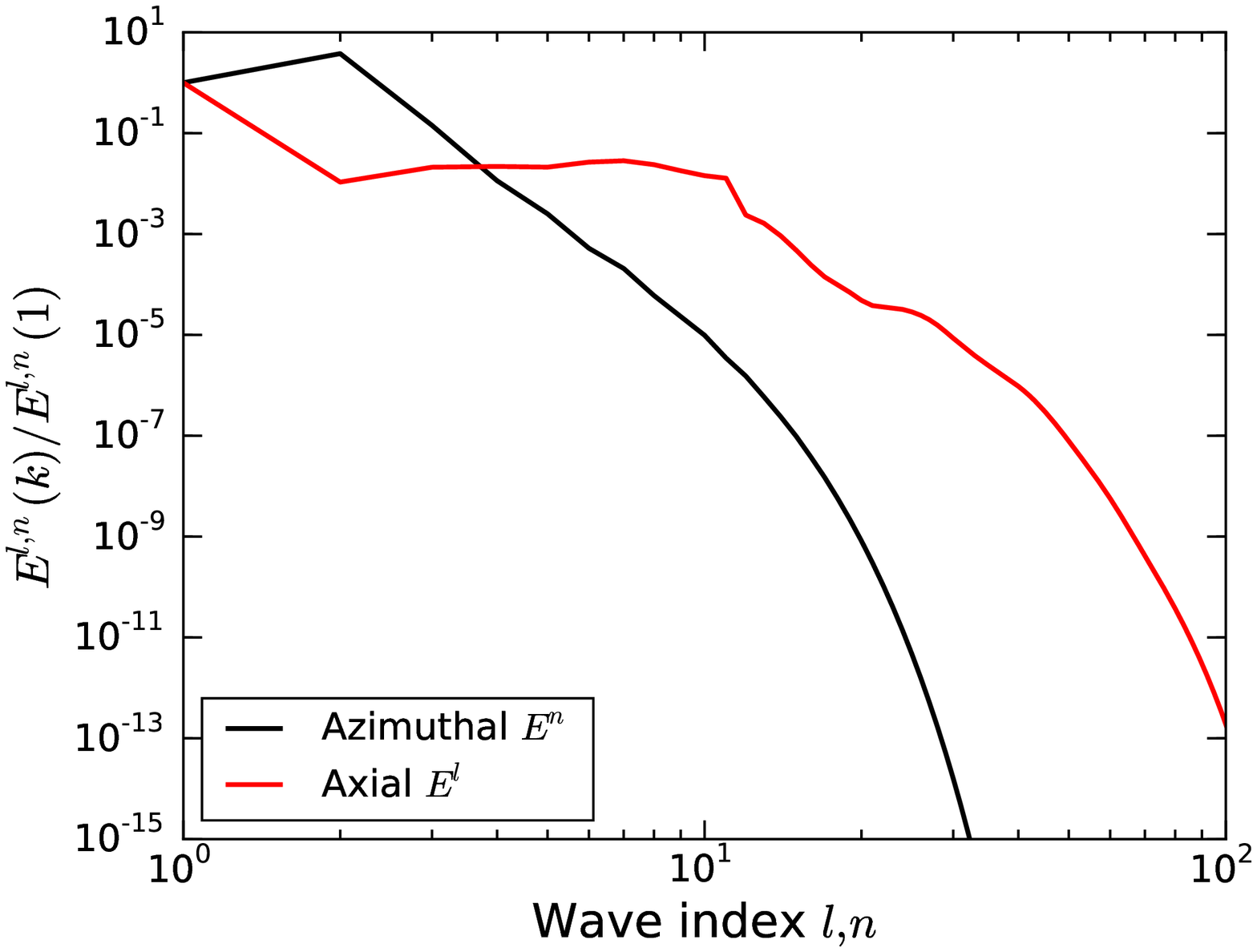}}
  \caption[Energy spectra $E^{l,n}(k)$]{
    Normalized axial (black) and azimuthal (red) energy power spectra $E^{l,n}(k)$ for (a) $Re_i=2\times 10^4$ 
    at time $t/\tau_d=30$ and for (b) $Re_i=2\times10^5$ at time $t/\tau_d=35$, before the decay of turbulence.   
    Here $l$ and $n$ are the wave indices in the axial and azimuthal directions, respectively. 
    The corresponding wavenumbers are $l\,k_z$ and $n\,k_{\theta}$.
    }
  \label{fig:eneSpe}
\end{figure}

For the simulations we employ our parallel code {\em nsCouette}~\cite{ShiAvila_caf2015}
which uses a spectral Fourier--Galerkin method for the discretization of the
Navier--Stokes equation
in the axial and azimuthal directions, and high-order finite differences in the 
radial direction, together with a second-order, semi-implicit projection scheme
for the time integration, and employs a pseudospectral method for the evaluation of the
nonlinear terms.
The corresponding parameters of the simulations are listed in table~\ref{tab:numParTCF}. At $Re_i=10^4$ we simulate a quarter of the cylinder in the azimuthal direction, corresponding to a basic azimuthal wavenumber $k_\theta=4/r_\text{mid}$, with $r_{mid} = 0.5(1+\eta)/(1-\eta)$, and set $\Gamma=L_z/d=0.5$, corresponding to a basic axial wavenumber $k_z=4\pi$. The total number of grid points before de-aliasing is $(N_r \times N_{\theta} \times N_z)$ is $(256 \times 512 \times 256)$. To save computing time, the domain size at higher Reynolds numbers is chosen smaller, $k_\theta=8$ and $16$ (the factor $1/r_{\text{mid}}$ is hereafter omitted for simplicity), at $Re_i=10^5$ and $Re_i=2\times10^5$, respectively. As shown in~\cite{BrauckmannEckhardt_jfm2012,Monico_pof2015}, a reduction of the domain length in the azimuthal direction has little effect on the statistical properties of the simulated turbulent flows, as long as the dominant structures are still captured. At high $Re$ the spatial resolution in each direction is increased approximately as $N\sim Re^{3/4}$, given that the domain size is the same. The resolution is checked at $Re_i=2\times 10^4$ and $Re_i = 2\times 10^5$ by the axial and azimuthal energy spectra as a function of the wave index $l,n$ (see Fig.~\ref{fig:eneSpe}). We should point out that in the case III a lower resolution than the one shown in Table~\ref{tab:numParTCF} causes the simulations to blow up. This may be explained by the fact that with a low resolution the scales at which energy dissipates is not resolved so that the energy accumulates in the flow and causes the simulations to diverge. 

\begin{table} [!h]
  \centering
  \begin{tabular}{c c c c c c c}
    \hline \hline 
    No. & $Re_i$ & $Re_s$ & $k_{\theta}$ & \# points & $dt/\tau_{d}$ \\
    \hline 
    I & $1\times 10^4$ & 5078.8 & $4$ & $256\times 512 \times 256$ & $10^{-5}$ \\
    II & $2\times 10^4$ & 10157.6 & $8$ & $256\times 512 \times 256$ & $2\times 10^{-5}$ \\
    III & $1\times 10^5$ & 50788 & $16$ & $1152\times 384 \times 384$ & $10^{-4}$ \\
    IV & $2\times 10^5$ & 101576 & $16$ & $2048\times 768 \times 512$ & $10^{-4}$ \\
    \hline \hline
  \end{tabular}
  \caption[Numerical parameters in TCf simulations]{DNS parameters of TCf in the quasi-Keplerian regime. 
    The radius ratio is $\eta = 0.71$ and the length-to-gap aspect ratio in the axially periodic direction is $\Gamma = 0.5$.}
  \label{tab:numParTCF}
\end{table}
\pagebreak
Our initial conditions are optimal perturbations from the computations of the transient growth by Maretzke {\em et al.}~\cite{Maretzke_jfm2014}, on top of which small three dimensional random noise exciting axial modes $l=1,\cdots,10$ is added. The optimal perturbations are computed at fixed $k_{\theta}$ (e.g.~$k_\theta=4$ at $Re_s=10^4$) and hence are optimal only in their subspace. Using the full domain in the azimuthal direction ($k_\theta=1$) would yield slightly higher transient growth. In all cases the azimuthal wavenumber of the optimal perturbation is chosen to be the same as the basic azimuthal wavenumber fixing the domain length in the azimuthal direction. For the case $R_{\Omega}=-1.2$ and $Re_s>\mathcal{O}(10^3)$, the optimal axial wavenumber is $k_z=0$, corresponding to an axially-invariant Taylor-column-like structure. In the $r-\theta$ plane, the optimal perturbation has an elongated spiral structure, similar to Fig.~8(a) in~\cite{Maretzke_jfm2014}, and extracts energy from the basic flow via the Orr mechanism. The optimal transient growth energy values, denoted as $G^{\text{opt}}$ (the mathematical definition can be found in~\cite{Maretzke_jfm2014}), at the investigated Reynolds numbers are listed in table~\ref{tab:transGrowth}. 

The initial velocity field $\textbf{u}_0$ is composed of three parts: the base flow $\textbf{U}_b$, the 2D optimal perturbation $\textbf{u}_0^{2D}$ and the 3D noise $\textbf{u}_0^{3D}$: $\textbf{u}_0=\textbf{U}_b+\textbf{u}_0^{2D}+\textbf{u}_0^{3D}$. The relative magnitude of the amplitude of the three components is $||\textbf{U}_b|| \gg ||\textbf{u}_0^{2D}|| \gg ||\textbf{u}_0^{3D}||$. All simulations were performed on standard HPC clusters with Intel processors and InfiniBand interconnect. The simulations are computationally expensive: simulation IV, for example, was performed on the high-performance system \emph{Hydra} at the Max Planck Computing and Data Facility and required about $5\times 10^6$ core hours using 5120 cores utilized by 512 MPI tasks (2 tasks per 20-core-node) with 10 OpenMP threads each. 

\begin{table} [!h]
  \centering
  \begin{tabular}{c c c c c c c}
    \hline \hline 
    No. & $Re_i$  & $k_z$ & $k_{\theta}$ & $G^{\text{opt}}$ & $t^{\text{opt}}/\tau_{d}$\\
    \hline 
    I & $1\times 10^4$ & 0 & 4 & 13.04  & 27 \\
    II & $2\times 10^4$ & 0 & 8 & 24.40  & 22 \\
    III & $1\times 10^5$ & 0 & 8 & 73.98 & 36 \\
    IV & $2\times 10^5$ & 0 & 16 & 82.13 & 28 \\
    \hline \hline
  \end{tabular}
  \caption[Transient growth rates of the optimal perturbations]{
    The transient growth rate of the initial perturbations attained at time $t^{\text{opt}}$ 
    and their corresponding wavenumbers.}
  \label{tab:transGrowth}
\end{table}

We use the total perturbation kinetic energy as a diagnostic quantity. Assuming that $\hat{\textbf{u}}^{ln}(r) = \hat{\textbf{u}}(r,lk_{\theta},nk_z)$ are the spectral coefficients in Fourier space of the velocity field $\textbf{u}(r,\theta,z)$, the modal kinetic energy density $E^{ln}$ associated with the Fourier 
mode $(l,n)$ is defined as
\begin{equation}
  \centering
    E^{ln}=\frac{1}{2}\int_{r_i}^{r_o}[\hat{u}_r^{ln}(r)^2+\hat{u}_{\theta}^{ln}(r)^2+\hat{u}_z^{ln}(r)^2]rdr.
    \label{eq:Eln}
\end{equation}
We also analyze the contributions of the kinetic energy of the axial mode $l$ and of the azimuthal mode $n$, respectively,
\begin{equation}
  \centering
  E^l = \sum_{n=-N}^{N}E^{ln}, \quad E^n = \sum_{l=-L}^{L}E^{ln}.
  \label{eq:ElEn}
\end{equation}
The total kinetic energy can therefore be expressed as 
\begin{equation}
  \centering
  E = \sum_{l=-L}^{L}\sum_{n=-N}^{N}E^{ln}.
  \label{eq:E}
\end{equation}
By removing the laminar part from the total energy 
we obtain the perturbation energy $E_p$, 
which is defined according to Eq.~\ref{eq:Eln} but 
replacing $\hat{u}_{\theta}^{00}(r)$ with $[\hat{u}_{\theta}^{00}(r)-U_{\theta}^b(r)]$. 
Note that the spectral coefficient at $l=0$ and $n=0$, $\hat{u}_{\theta}^{00}(r)$, is 
the average azimuthal velocity.

\section{Results}

\subsection{Nonlinear transient growth}

\begin{figure}[!ht]
  \centering
  \vspace{3em}
  \includegraphics[width=0.45\textwidth]{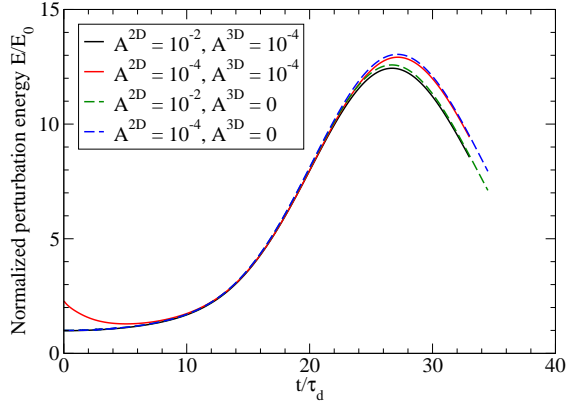}
  \caption{The nonlinear evolution of the perturbation kinetic energy normalized by 
    the initial energy of the 2D perturbation and 3D noise at $Re_i=10^4$. 
    Dashed lines: Without 3D noise, i.e., $A^{3D}=0$;
    Solid lines: With 3D noise, excited from $k_z=1,\cdots,10$. 
    The relative amplitude of the 3D noise is $A^{3D}=10^{-4}$, normalized by the inner Reynolds number $Re_i$.
    Colors indicate different relative 2D amplitudes.}
  \label{fig:ke1e4}
\end{figure}

The behavior of the transient growth of the initial 2D optimal perturbation at $Re_i=10^4$ and $k_\theta=4$ is first investigated. Two groups of simulations were performed: with and without 3D noise. Let $A^{2D}$ and $A^{3D}$ denote the relative amplitude of 2D perturbation and 3D noise, scaled by the inner Reynolds number $Re_i$ (the circular Couette flow). $A^{2D}=10^{-4}$ means that the absolute amplitude of the 2D perturbation is $10^{-4}\times Re_i = 1$. In order to test the effect of nonlinear terms, two runs with different relative 2D amplitude $A^{2D}=10^{-4}, 10^{-2}$ and without noise have been conducted. The time evolution of the perturbation kinetic energy normalized by the initial value is shown in Fig.~\ref{fig:ke1e4} (dashed lines). At low perturbation amplitude $A^{2D}=10^{-4}$ the maximum amplification $G^{\text{opt}}=13.04$ is attained at $t/\tau_{d}=27$, in excellent agreement with the linear prediction (see Table~\ref{tab:transGrowth}). With an amplitude $A^{2D}=10^{-2}$, the transient growth rate is slightly reduced due to the non-negligible nonlinear effects. When adding noise similar transient growth behavior is found, see Fig.~\ref{fig:ke1e4} (solid lines). Because of the negligible nonlinear effect, the added 3D noise decays monotonically to zero and seems to have no influence on the dynamics of 2D optimal perturbations. 

\subsection{Transition and decay of turbulence}

\begin{figure}[!ht]
  \centering
  \vspace{3em}
  \subfloat[$Re_i=2\times 10^4$]{\includegraphics[width=0.45\textwidth]{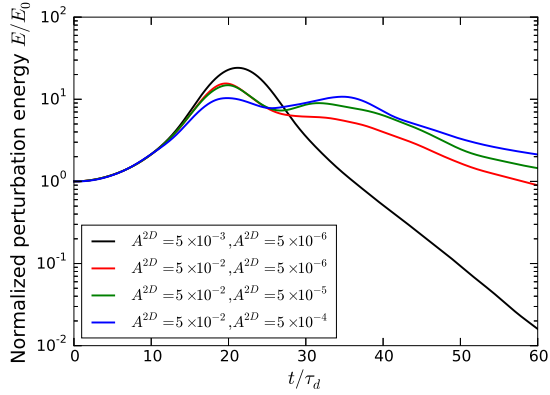}}\\[3em]
  \subfloat[$Re_i=2\times 10^5$]{\includegraphics[width=0.45\textwidth]{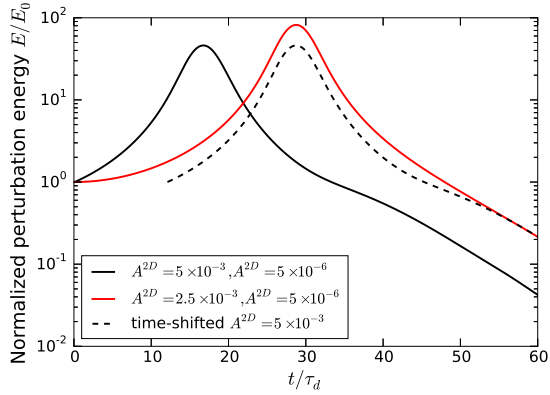}}
  \caption{The temporal evolution of the normalized perturbation kinetic energy at 
    $Re_i=2\times 10^4$ (a) and $Re_i=2\times 10^5$ (b). The line colors correspond to different perturbation amplitudes.
    \added{The dashed line is the same as the black solid line, but with a time shift of $12 \tau_d$.}}
  \label{fig:ke2e4}
\end{figure}

\begin{figure}[!ht]
  \centering
  \subfloat[$A^{2D}=5\times 10^{-3}, A^{3D}=5\times 10^{-6}$]{\includegraphics[width=0.45\textwidth]{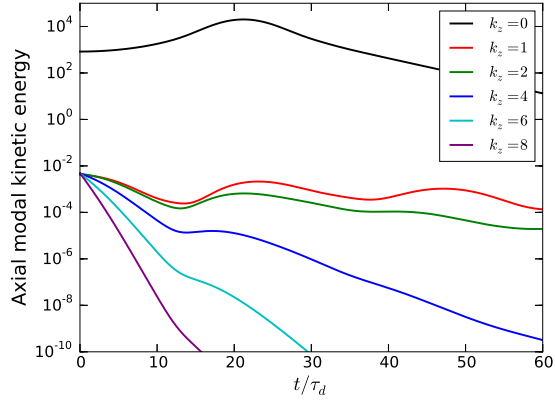}}\\[3em]
  \subfloat[$A^{2D}=5\times 10^{-2}, A^{3D}=5\times 10^{-6}$]{\includegraphics[width=0.45\textwidth]{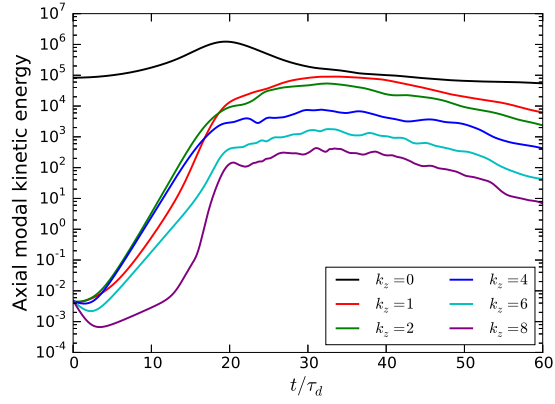}}
  \caption{The temporal evolution of the axial modal kinetic energy at 
    $Re_i=2\times 10^4$ for $A^{2D}=5\times 10^{-3}, A^{3D}=5\times 10^{-6}$ (a), and 
    $A^{2D}=5\times 10^{-2}, A^{3D}=5\times 10^{-6}$ (b).The different lines 
    correspond to different axial modes as indicated in the legend.}
  \label{fig:kz2e4}
\end{figure}

By increasing the amplitude of the 2D perturbations or 3D noise above a certain level, nonlinear effects become important and qualitatively change the dynamics of the flow. This has been observed at all Reynolds numbers investigated, and we first focus on the results at $Re_i=2\times 10^4$ using $k_\theta=8$, for which the transient growth of the 2D optimal perturbation is $G^{\text{opt}}=24.4$, attained at $t/\tau_{d}=22$. Here, four runs have been performed, based on different relative amplitudes of 2D perturbations and 3D noise: 
\begin{enumerate}
\item $A^{2D}=5\times 10^{-3}, A^{3D}=5\times 10^{-6}$
\item $A^{2D}=5\times 10^{-2}, A^{3D}=5\times 10^{-6}$
\item $A^{2D}=5\times 10^{-2}, A^{3D}=5\times 10^{-5}$
\item $A^{2D}=5\times 10^{-2}, A^{3D}=5\times 10^{-4}$ 
\end{enumerate}
The temporal evolution of the normalized perturbation kinetic energy for all these cases is shown in the top panel of Fig.~\ref{fig:ke2e4}. Interestingly, the flow dynamics for $A^{2D}=5\times 10^{-3}$ and $A^{2D}=5\times 10^{-2}$ are qualitatively different. At lower amplitude $A^{2D}=5\times 10^{-3}$, the flow cosely follows the path of the linear transient growth, with a maximum of about 24.15 at  $t/\tau_d=22$, followed by an exponential decay. However, at $A^{2D}=5\times 10^{-2}$, a second ``peak'' or ``bump'' appears after the initial transient growth, especially for the cases with larger 3D noise. In addition, the transient growth is reduced and occurs earlier if the level of 3D noise is large. The reason behind this qualitatively different behaviour at $A^{2D}=5\times 10^{-3}$ and $A^{2D}=5\times 10^{-2}$ is apparent in Fig.~\ref{fig:kz2e4}, where the axial modal kinetic energy $E^n(t)$ is shown. In both cases, the mode $k_z=0$ shows the initial transient growth as predicted by linear analysis. However, at the initial stage, the higher axial energy modes for $A^{2D}=5\times 10^{-2}$ experience exponential or even faster growth, whereas for $A^{2D}=5\times 10^{-3}$ they all decay. 

\begin{figure}[!ht]
  \centering
  \vspace{3em}
  \subfloat[$A^{2D}=2.5\times 10^{-3}, A^{3D}=5\times 10^{-6}$]{\includegraphics[width=0.45\textwidth]{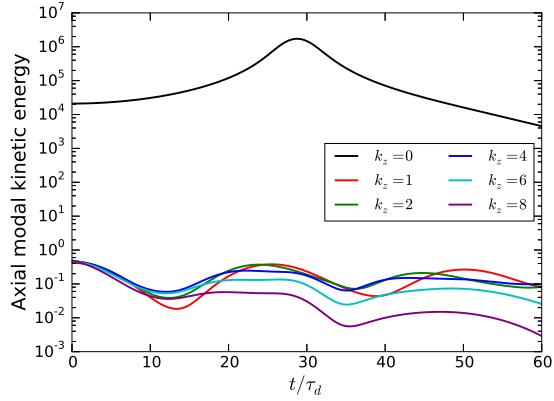}}\\[3em]
  \subfloat[$A^{2D}=5\times 10^{-3}, A^{3D}=5\times 10^{-6}$]{\includegraphics[width=0.45\textwidth]{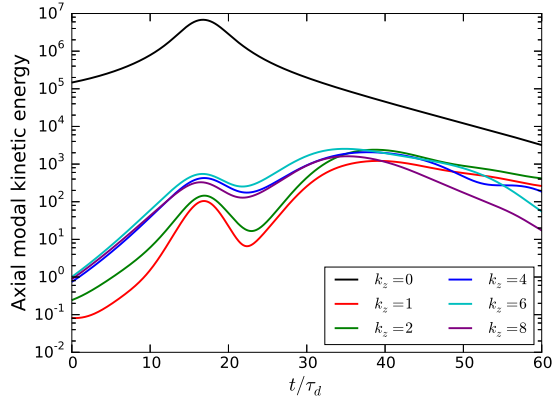}}
  \caption{The temporal evolution of the axial modal kinetic energy at 
    $Re_i=2\times 10^5$ for $A^{2D}=2.5\times 10^{-3}, A^{3D}=5\times 10^{-6}$ (a) and 
    $A^{2D}=5\times 10^{-3}, A^{3D}=5\times 10^{-6}$ (b). Different colors correspond to different axial modes.}
  \label{fig:kz2e5}
\end{figure}

At $Re_i=2\times10^5$ and $k_\theta=16$ simulations were done for $A^{2D}=2.5\times 10^{-3}$ and $A^{2D}=5\times 10^{-3}$. The temporal evolution of the normalized perturbation kinetic energy is shown in \removed{the bottom panel of }Fig.~\ref{fig:ke2e4}\added{b}. \changed{At higher amplitude, the initial transient growth peak occurs at an earlier moment and its value is lower than the linear prediction, while at the lower amplitude the perturbation energy follows basically the linear dynamics (the same behavior is observed for $Re_i=10^5$). When we look at the evolution of the axial modal energy (Fig.~\ref{fig:kz2e5}), the difference is obvious, as has been found at $Re_i=2\times 10^4$. At higher perturbation amplitude, the exponential growth of higher axial modes drives the flow temporarily chaotic}{For low initial amplitude, the perturbation energy follows closely the linear dynamics, whereas at high initial amplitude nonlinear effects become important, as observed at $Re_i=2\times10^4$. The effect of nonlinearity is to reduce the energy amplification and in addition the peak energy is reached here much earlier (by approximately $12\tau_d$). However, the two temporal evolutions are qualitatively similar and can be collapsed together by shifting the curve for $A^{2D}=2.5\times 10^{-3}$ by $12\tau_d$ horizontally and then vertically so that they have the same amplitude at $t=12\tau_d$ (see the dashed curve in  Fig.~\ref{fig:ke2e4}b). It thus appears that the effect of nonlinearity is essentially to accelerate the initial phase of the disturbance evolution. This reduces the maximum energy growth, but not very substantially because in the initial phase the optimal mode is  weakly amplified. The lion's share of the energy amplfication occurs as the vortices are tilted by the shear (Orr mechanism) and change their orientation angle \cite{Maretzke_jfm2014}, which occurs in both cases. }

\added{The axial modal energies behave qualitatively differently depending on the initial perturbation amplitude (see Fig.~\ref{fig:kz2e4} for $Re_i=2\times10^4$ and Fig.~\ref{fig:kz2e5} for $Re_i=2\times10^5$). At low amplitude the axial modes oscillate in time while being damped, whereas at thigh amplitude the modified velocity profile is linearly unstable at $t=0$ and the leading axial modes grow exponentially, as expected in a secondary instability. The flow turns temporarily chaotic,} but the ensuing turbulent motions finally decay and the flow returns to laminar. At $Re_i=2\times 10^5$, the modal energy is much higher than $Re_i=2\times 10^5$ because of stronger nonlinear interactions and the relaminarization process, which is controlled by viscosity, takes much longer  when measured in advective time units. In summary, the following conclusions can be drawn:

\begin{enumerate}
\item At small perturbation amplitude nonlinear effects are negligible and 
    the flow follows the linear dynamics.
\item At large enough perturbation amplitude, the initial maximum growth of the total energy is smaller and attained at an earlier moment.
\item Transition to turbulence occurs via three-dimensional secondary instabilities of the flow modified by the optimal disturbance~\cite{SchmidHenningson_springer2001}.
\item The resulting hydrodynamic turbulence at $Re_s$ up to $10^5$ is not sustained and eventually decays.
\end{enumerate}
\pagebreak

\begin{figure*}[!ht]
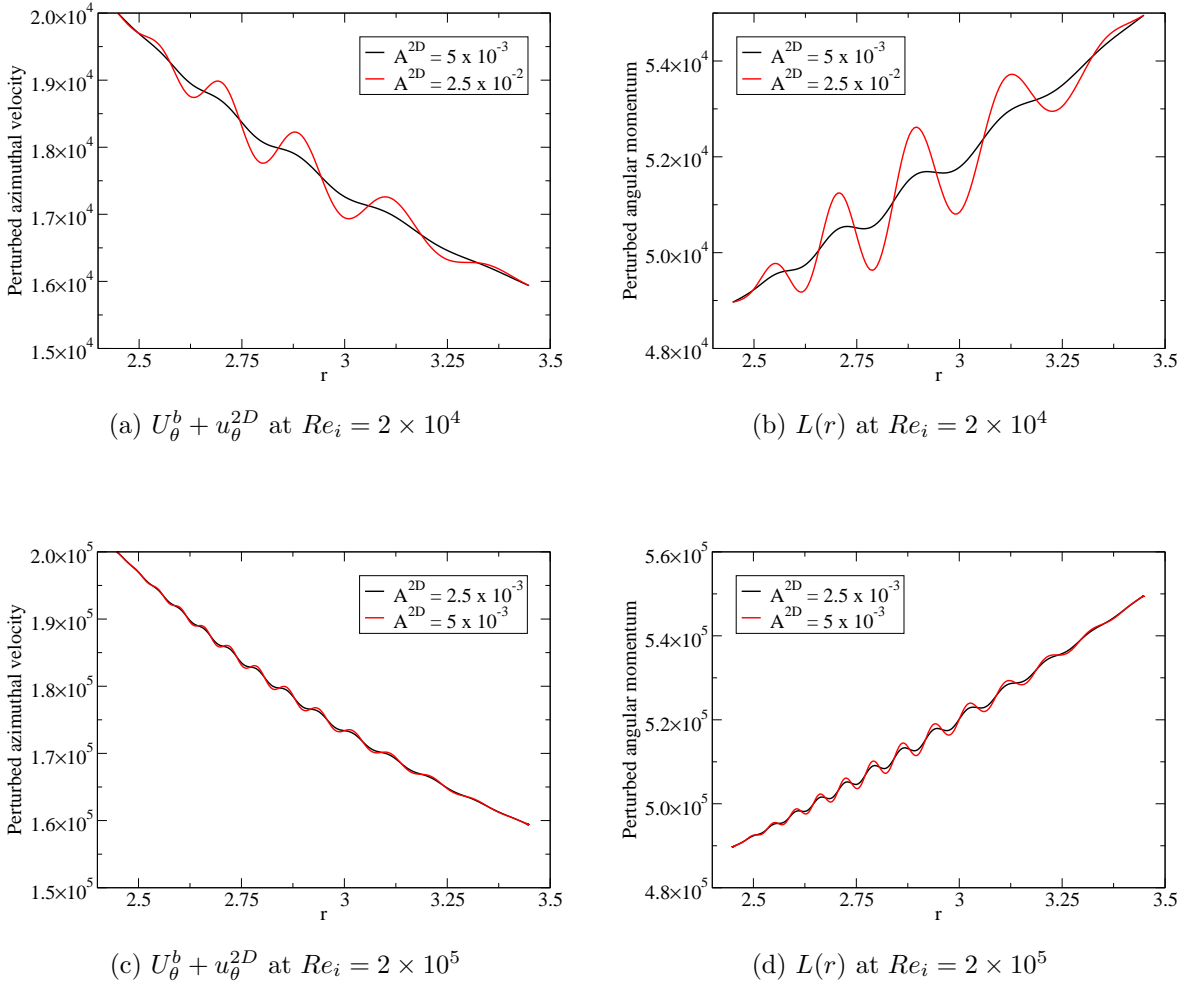

  \centering
  \vspace{3em}
  \subfloat[$U_{\theta}^b + u_{\theta}^{2D}$ at $Re_i=2\times 10^4$]{\includegraphics[width=0.45\textwidth]{perturbUbase_2e4.eps}}\qquad
  \subfloat[$L(r)$ at $Re_i=2\times 10^4$]{\includegraphics[width=0.45\textwidth]{angularM_2e4_5e2.eps}}\\[3em]
  \subfloat[$U_{\theta}^b + u_{\theta}^{2D}$ at $Re_i=2\times 10^5$]{\includegraphics[width=0.45\textwidth]{perturbUbase_2e5.eps}}\qquad
  \subfloat[$L(r)$ at $Re_i=2\times 10^5$]{\includegraphics[width=0.45\textwidth]{angularM_2e5.eps}}
  \caption{Radial profiles of the perturbed azimuthal velocity (a, c) and 
    Angular momentum $L(r)=(U_{\theta}^b + u_{\theta}^{2D})r$ (b, d) at $t=0$ for 
    $Re_i=2\times 10^4$ (a, b) and $Re_i=2\times 10^5$ (c, d). }
  \label{fig:pertUbase}
\end{figure*}

\begin{figure}[!ht]
  \centering
  \subfloat[$t/\tau_d=4$]{\includegraphics[width=0.45\textwidth]{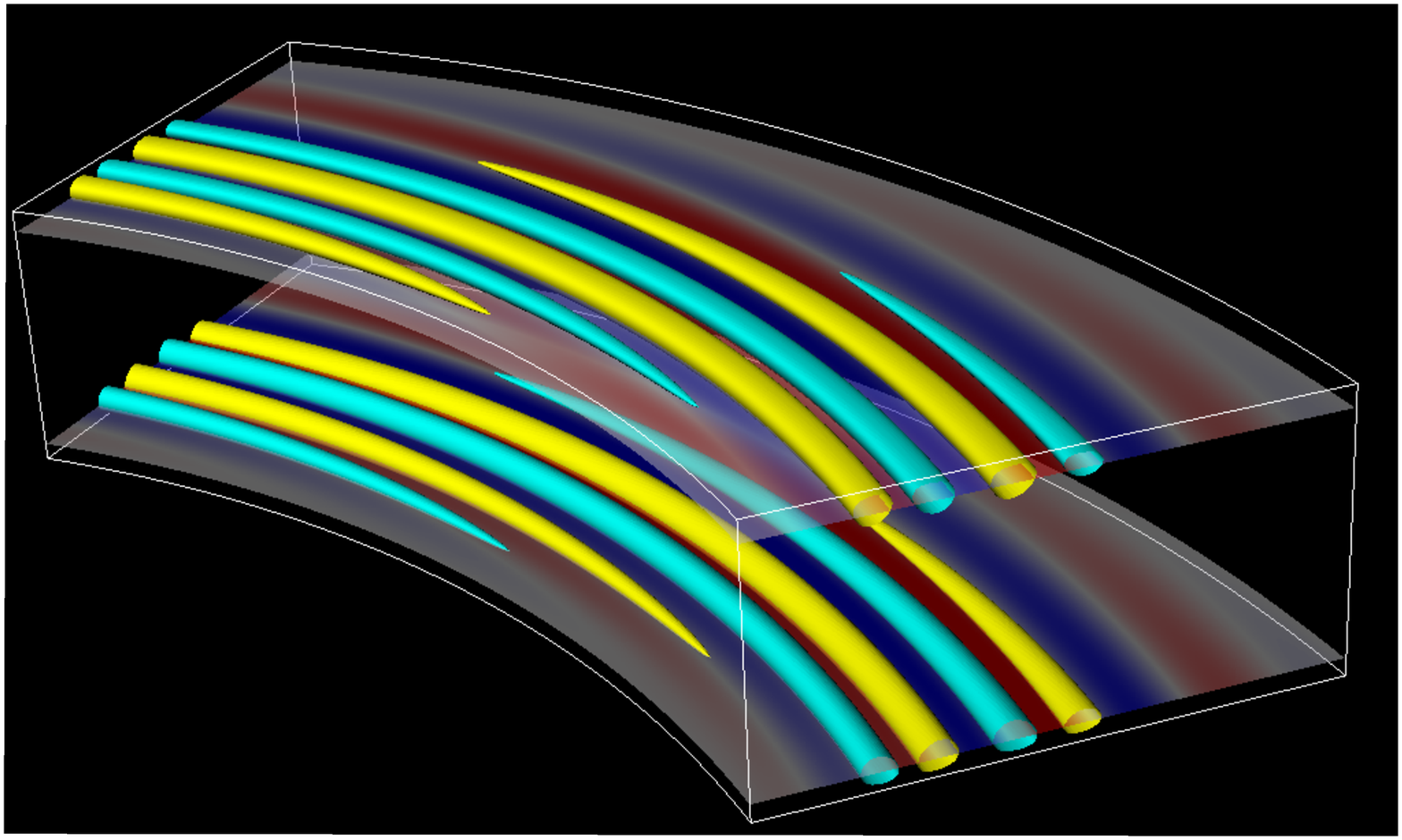}} \qquad
  \subfloat[$t/\tau_d=8$]{\includegraphics[width=0.45\textwidth]{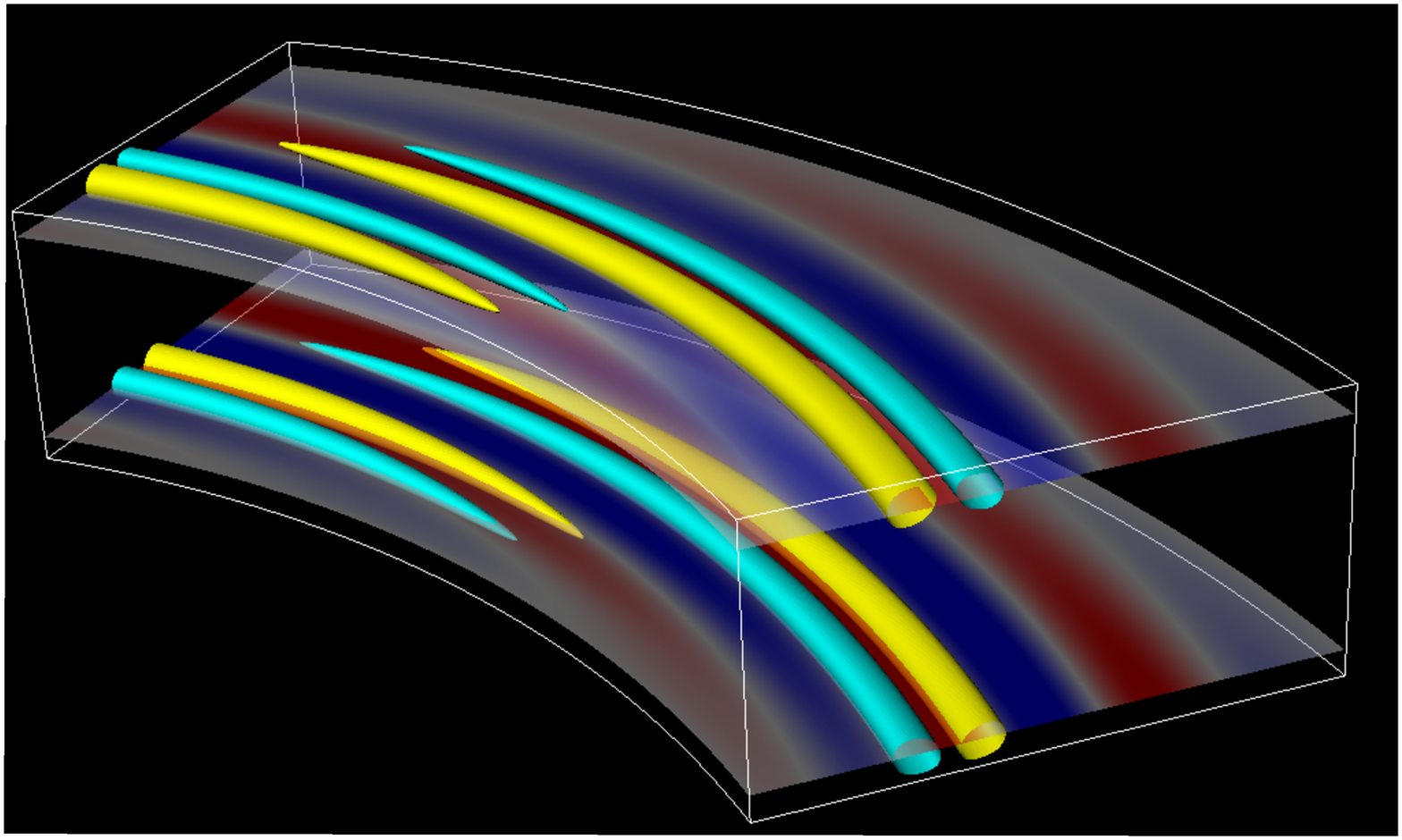}}\\
  \subfloat[$t/\tau_d=12$]{\includegraphics[width=0.45\textwidth]{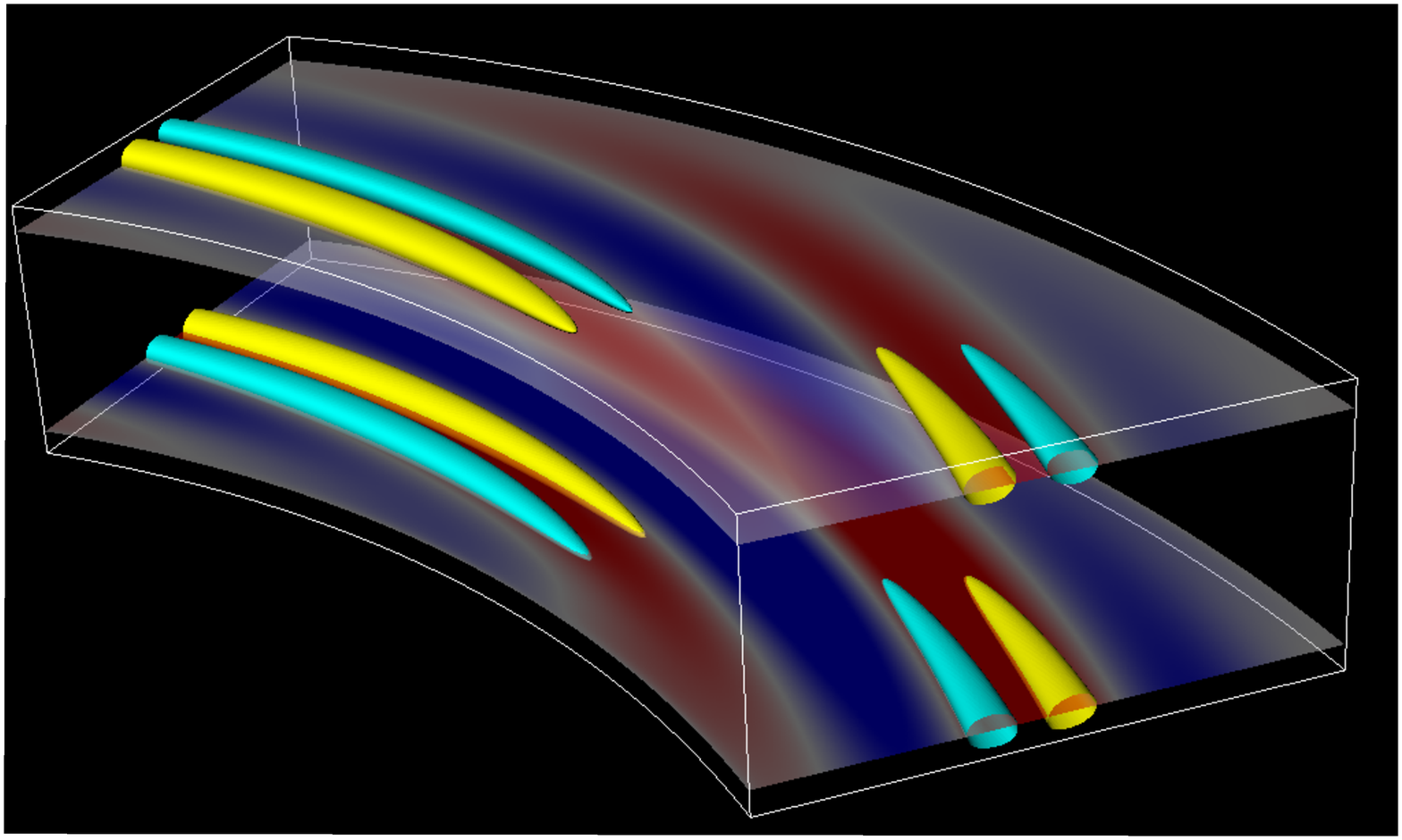}}\qquad
  \subfloat[$t/\tau_d=16$]{\includegraphics[width=0.45\textwidth]{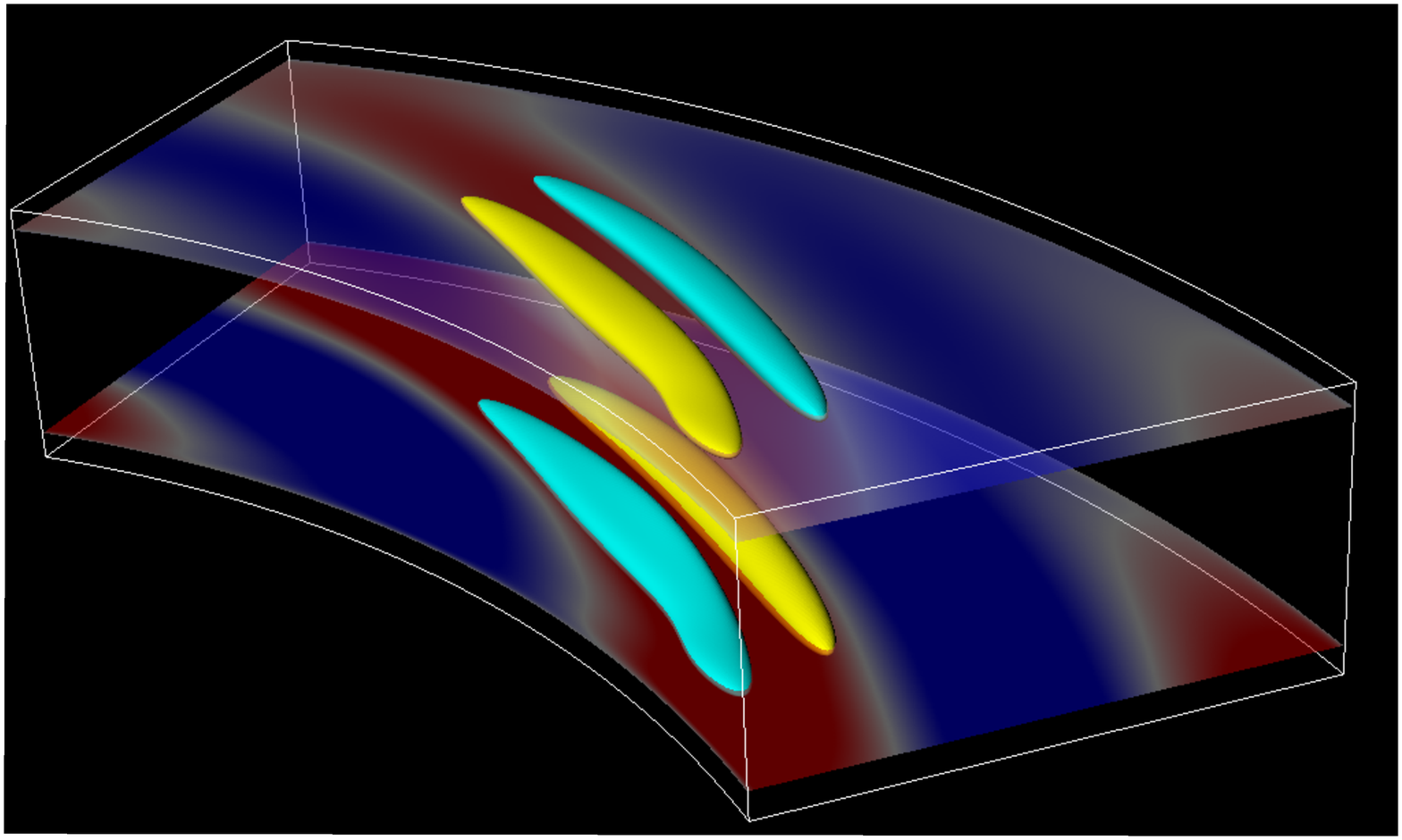}}
  \caption{Isosurfaces of the instantaneous vertical velocity $u_z$ at $Re_i=2\times 10^4$ for $A^{2D}=2.5\times 10^{-2}$ and $A^{3D}=5\times 10^{-6}$ at time $t/\tau_{d}=4$ (a), $t/\tau_{d}=8$ (b), $t/\tau_{d}=12$ (c) and $t/\tau_{d}=16$ (d). 
    Two iso-levels are used: Yellow indicates positive $u_z$ and cyan is for negative 
    $u_z$. The two horizontal planes show the radial derivative of the angular momentum $dL/dr$. The (symmetric) colour scale varies from red (positive) over white (zero) to blue (negative).}
  \label{fig:uz2e4}
\end{figure}

\begin{figure*}[!ht]
  \centering
  \subfloat[$t/\tau_d=25$]{\includegraphics[width=0.4\textwidth]{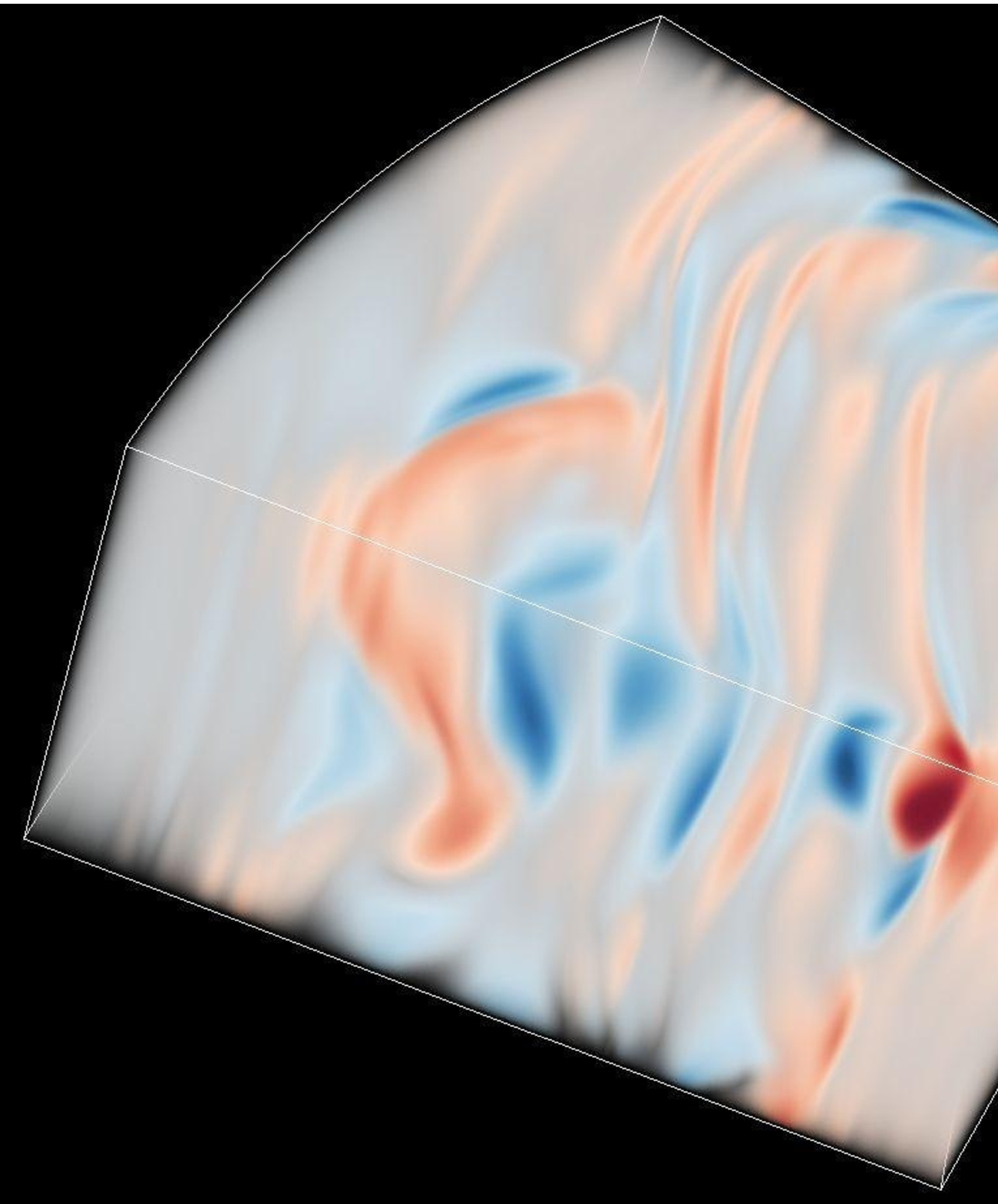}}\qquad
  \subfloat[$t/\tau_d=40$]{\includegraphics[width=0.4\textwidth]{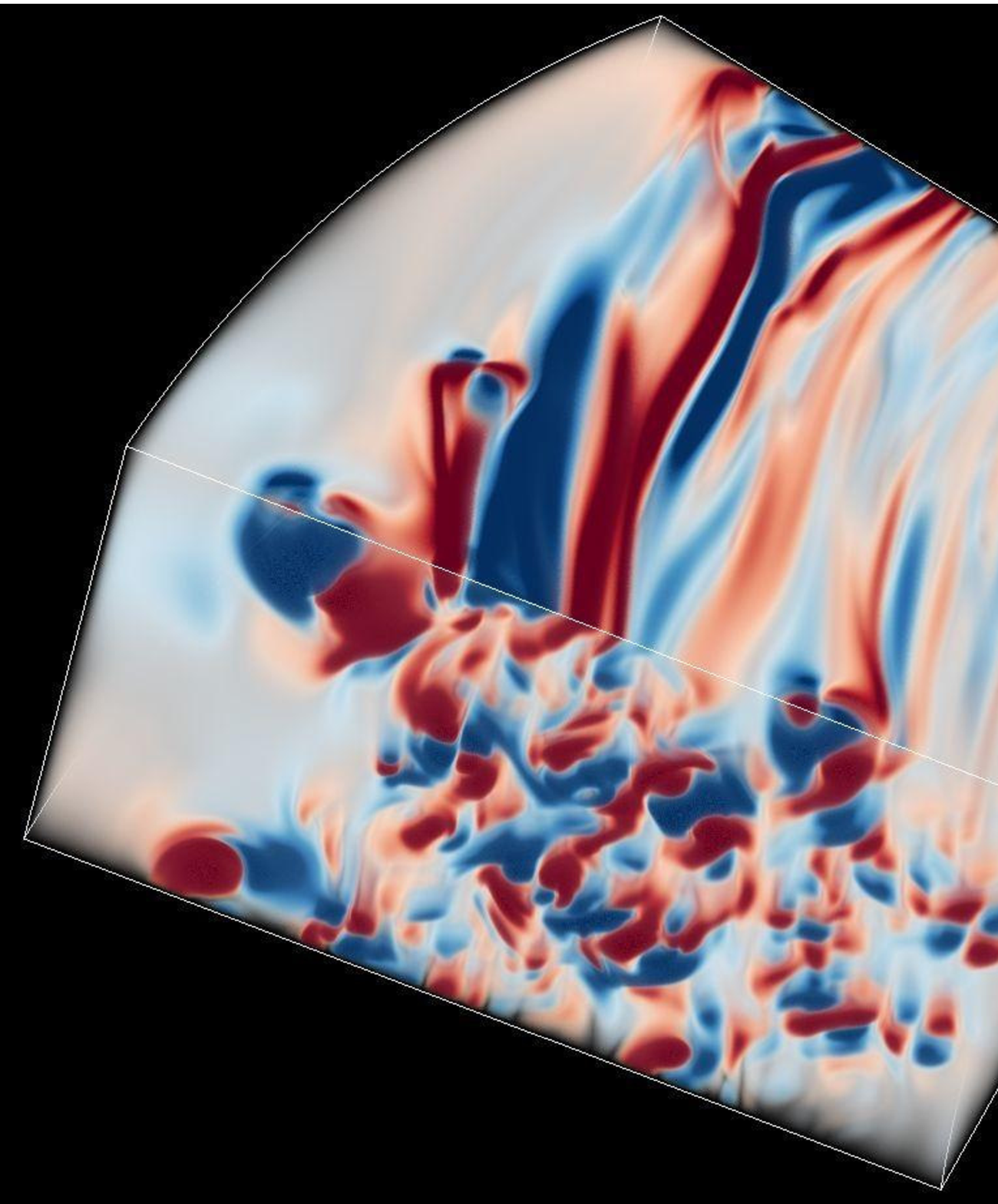}}\\
  \subfloat[$t/\tau_d=55$]{\includegraphics[width=0.4\textwidth]{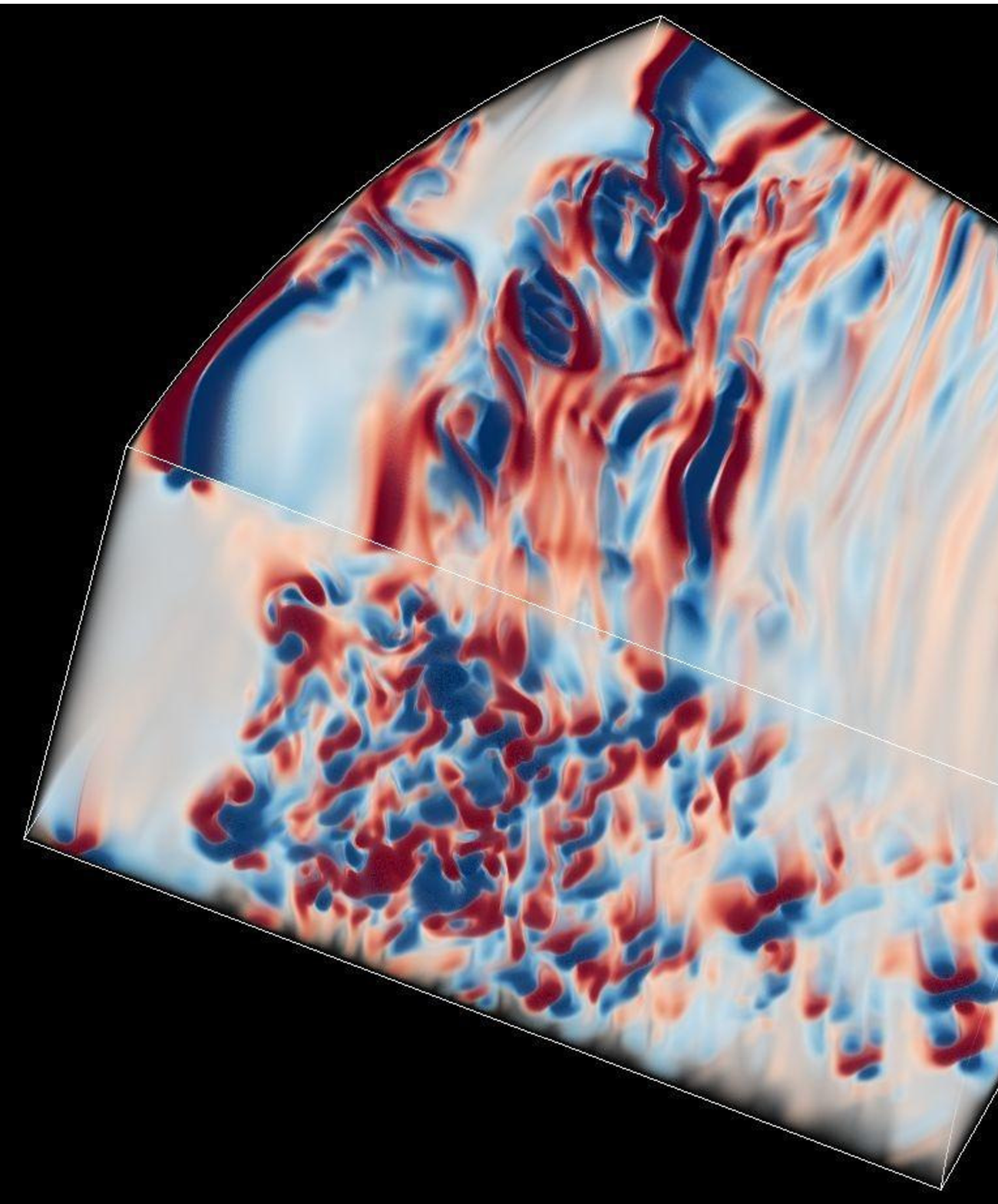}}\qquad
  \subfloat[$t/\tau_d=70$]{\includegraphics[width=0.4\textwidth]{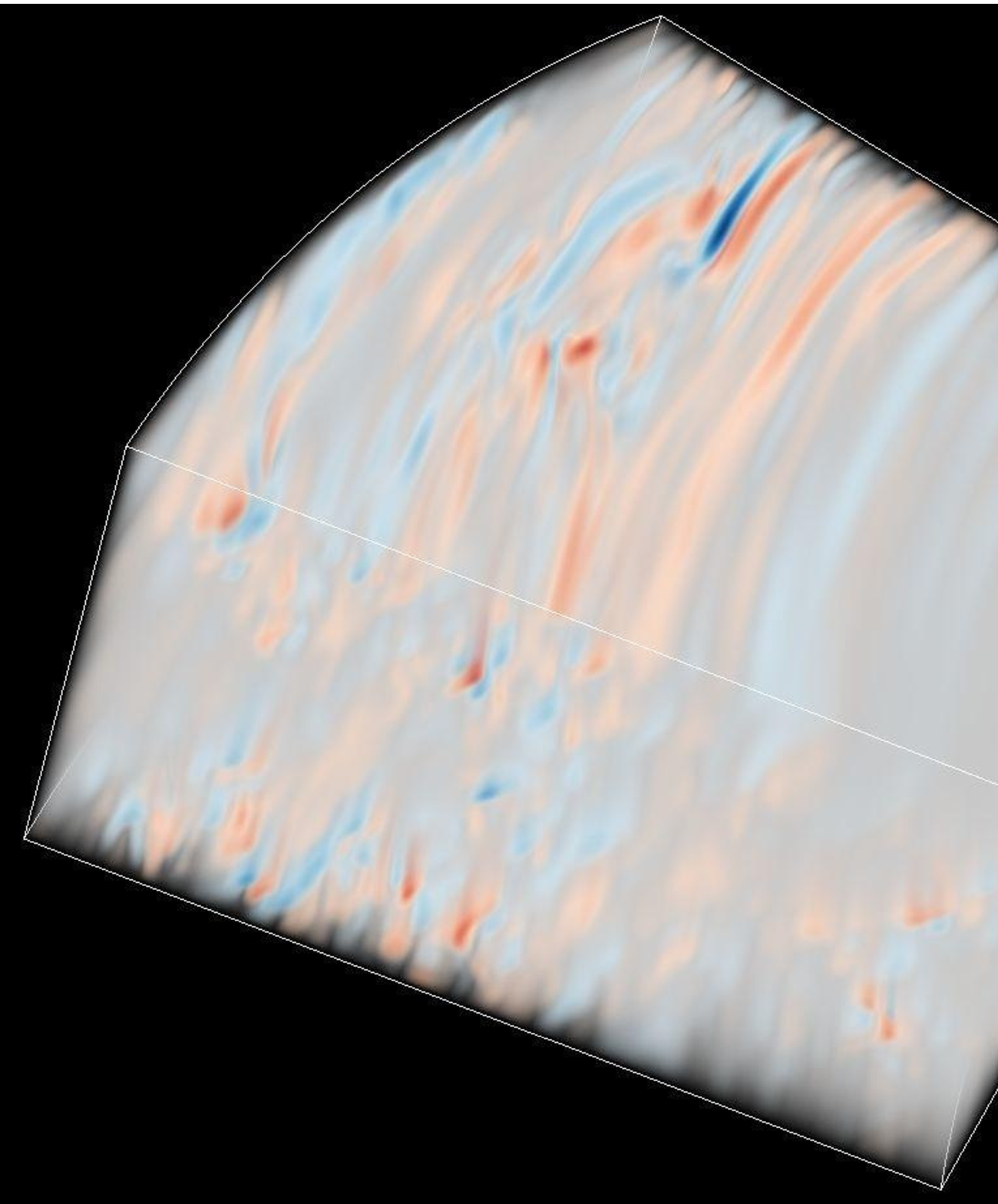}}
  \caption{Volume rendering of the instantaneous streamwise vorticity $\omega_{\theta}$ at $Re_i=2\times 10^5$ for $A^{2D}=5\times 10^{-3}$ and $A^{3D}=5\times 10^{-6}$ at times $t/\tau_{d}= 25$ (a), $t/\tau_{d}=40$ (b), $t/\tau_{d}=55$ (c) and $t/\tau_{d}=70$ (d). Red (blue) colours trace regions with positive (negative) $\omega_{\theta}$. (Multimedia view)}
  \label{fig:wth2e5}
\end{figure*}

A remaining intriguing issue concerns the physical mechanism responsible for the two distinct behaviors at different perturbation amplitude as described above. Dubrulle and Knobloch~\cite{DubrulleKnobloch_aa1992} proposed that finite amplitude perturbations may generate inflection points in the base profile, which cause secondary instabilities and breakdown to turbulence. A similar mechanism was suggested in pipe flow by Meseguer~\cite{Meseguer_pof2003}, who performed simulations with different 2D and 3D perturbation amplitudes and observed sustained transition to turbulence at sufficiently large amplitudes. However, as shown in Fig.~\ref{fig:pertUbase} (a, c), the perturbed azimuthal velocity profiles all have inflection points but some fail to generate turbulence. Moreover, 
one important difference with secondary instability as observed in non-rotating shear flows, such as channel, Couette and pipe flow is that the amplitude of the optimal mode needed to trigger the secondary instability is very high~\cite{DarbyshireMullin_JFM1995, Dauchot_pof1995, Eckhardt_ARFM2007}. 
Figures~\ref{fig:kz2e4} and~\ref{fig:kz2e5} show that in fact the energy of the three-dimensional modes starts growing already at $t=0$ and not when the transient growth peaks. It is thus very unlikely that the transient growth is responsible for the observed transition. Instead it appears that at $t=0$ the base flow is already sufficiently distorted so that the flow is already linearly unstable. Figure~\ref{fig:pertUbase} (b, d) shows the radial distribution of the angular momentum $L(r) = (U_{\theta}^b + u_{\theta}^{2D})r$ at $t=0$ for $Re_i=2\times 10^4$ and $2\times 10^5$. The black curves correspond to runs in which no secondary instability is observed, whereas the red curves correspond to unstable runs. In the latter there are several regions in the flow in which the angular momentum decreases outwards, locally, and thus these regions are centrifugally unstable according to the Rayleigh criterion for inviscid rotating fluids. Figure~\ref{fig:uz2e4} shows the instantaneous vertical velocity $u_z$ at $Re_i=2\times 10^4$ at four different instants of the time evolution. The horizontal planes show false-color plots of the radial derivatives of angular momentum $dL/dr$. 
There are regions in which the angular momentum decreases steeply, thereby suggesting that the instability is centrifugal in nature. The emerging streamwise vortices are nearly axisymmetric and are reminiscent of Taylor vortex flow.
Note that the Rayleigh criterion is inviscid and viscosity has a stabilising effect, so that locally Rayleigh-unstable regions are not sufficient for flow instability to occur in viscous flows. In a Rayleigh-unstable region of length $l$ the viscous (Laplacian) term in the Navier--Stokes equation implies that the stabilizing effect is proportional to $1/l^2$, so that the smaller $l$ is, the larger the stabilising effect is. Hence in very small Rayleigh-unstable regions the instabilities are strongly suppressed by viscosity. Decaying turbulence is clearly observed at $Re_i=2\times 10^5$, where the flow is  much more turbulent, as shown in the volume rendering of the streamwise vorticity in Fig.~\ref{fig:wth2e5} (Multimedia view).

\section{Conclusion}

We  performed direct numerical simulations of axially periodic TCf in the quasi-Keplerian regime by strongly disturbing the laminar Couette flow. No sustained turbulence was found at shear Reynolds numbers up to $\mathcal{O}(10^5)$, in agreement with previous experiments (see Ref.~\cite{Edlund_pre2015} and references therein) and direct numerical simulations~\cite{Monico_jfm2014} using turbulent initial conditions. We used linear optimal perturbations (axially invariant Taylor columns) superposed with small three-dimensional noise. Depending on the initial perturbation amplitude, the flow dynamics vary significantly. At small amplitudes, the flow follows the path of linear transient growth, whereas at large initial amplitude the initial growth is reduced and the peak of the transient growth occurs at earlier times because of non-negligible nonlinear effects. For sufficiently large amplitudes transition to turbulence can be triggered followed by rapid decay driven by  viscous effects. 

The transition scenario found here is qualitatively different from that in wall-bounded shear flows without rotation. 
In the latter optimal disturbances are stream-wise aligned vortices and 
when used as initial conditions they create velocity streaks, 
which render the flow linearly unstable and subsequently turbulent~\cite{Zikanov_pof1996,SchoppaHussain_jfm2002}. 
This streak instability and the generation of streaks via stream-wise vortices 
are the essential ingredients for the self-sustenance of turbulence in wall-bounded 
shear flows~\cite{Hamilton_JFM1995, Waleffe_pof1997}. 
Instead, in quasi-Keplerian TCf stream-wise vortices are unable to 
efficiently extract energy from Couette flow~\cite{Maretzke_jfm2014}, 
and so they cannot contribute to a self-sustaining process~\cite{Rincon_aa2007}. 
Here the optimal disturbances are axially invariant vortices, 
and their transient growth is substantially smaller than for stream-wise vortices 
in wall-bounded shear flows without rotation~\cite{Maretzke_jfm2014}. 
Our simulations indicate that these axially invariant disturbances 
cannot generate a secondary instability unless they are so large that they already initially, 
i.e.~without energy growth, modify regions of the Couette flow 
so that these become locally Rayleigh unstable. 
This instability is unable to recreate the axially invariant optimal modes and so turbulence decays immediately after transition. 
Whether hydrodynamic turbulence can be sustained at even higher Reynolds number 
requires further research. 

\section*{Acknowledgement}
L. Shi and B. Hof acknowledge research funding by Deutsche Forschungsgemeinschaft (DFG, Germany) under grant No. SFB963/1 (Project A8). Computing time was allocated through the PRACE DECI-10 project HYDRAD and by the Max Planck Computing and Data Facility (Garching, Germany). 

\newpage

\bibliography{./ref_fluidMech.bib}

\end{document}